\documentclass[11pt]{article}
\usepackage{latexsym}
\raggedright
\newcommand{\T}{{\rm Tr} \,}

\newcommand{\cH}{{\cal H}}

\newcommand{\1}{{\bf 1}}

\newcommand{\m}{{\rm av}}

\begin{document}
  \title{Concurrence and foliations induced by some 1-qubit channels }
  \author{Armin Uhlmann}
  \date{Institute
  for Theoretical Physics\\ University of Leipzig}
  \maketitle

\begin{abstract}
We start with a short introduction to the roof concept.
An elementary discussion of phase-damping channels shows the role of
anti-linear operators in representing their concurrences. A general
expression for some concurrences is derived. We apply it to
1-qubit channels of length two, getting the induced
foliations of the state space, the optimal decompositions,
and the entropy of a state with respect to these channels.
For amplitude-damping channels one obtains an expression
for the Holevo capacity allowing for easy numerical calculations.
\end{abstract}

\section{Introduction}
The aim of this paper is in calculating some entanglement quantifying
functions for a rather restricted class of quantum channels, in
particular for all 1-qubit channels of length two.
These channels are now well examined and classified due to
the work of Fujiwara and Algoet \cite{FA98}, King and Ruskai
\cite{KR01}, Ruskai et al \cite{RSW00}, Verstraete and Verschelde
\cite{VV02}. They include all extremal 1-qubit channels and
some important doubly stochastic ones. An introduction to
quantum channels is in Nielsen and Chuang, \cite{NC00}.

A channel, say $T$, is of rank two, if $T(\rho)$ is of rank two
for all density operators. In this case there are at most two
eigenvalues of $T(\rho)$ different from zero. It follows from
trace preserving, that $T(\rho)$ is characterized by its largest
eigenvalue up to a general unitary transformation. Nevertheless,
the treatment of a general rank two channel is beyond our
present abilities if one asks for quantities like capacity or
concurrence.

A completely positive map is of length two, if it can be written
down with two Kraus operators, but not with one.

Let us now repeat some definitions and properties
of the quantities to be treated.
An {\em ensemble} is a finite number of density
operators, every one given with a definite probability or weight,
\begin{equation} \label{u11}
{\cal E} = \{ \rho_1, \dots, \rho_m ; p_1, \dots, p_m \} ,
\end{equation}
such that the sum of the positive numbers $p_j$ is one. The density
operators are {\em states} of a physical system.
We may think of a
{\em quantum alphabet} with {\em quantum letters} $\rho_j$,
and  of a {\em quantum message} which chooses randomly the
quantum letter $\rho_j$ with probability $p_j$.

The weighted sum of (\ref{u11}) is the {\rm average density} of
the ensemble,
\begin{equation} \label{u12}
\m[{\cal E}]  := \sum p_j \rho_j
\end{equation}
While the quantum letters of ${\cal E}$ are states of a physical
system, the average (\ref{u12}) is a ``non-commutative probability
measure''. Its role is quite similar to a ``classical''
discrete probability distribution.
The extractable  information  per quantum letter
of the quantum message is bounded by Holevo's \cite{Ho73}
\begin{equation} \label{u13}
\chi({\cal E}) := S(\rho) - \sum p_j S(\rho_j)
= \sum p_j S( \rho \parallel \rho_j ), \quad \rho =  \m[{\cal E}]
\end{equation}
Here $S(.)$ is the von Neumann entropy and $S(. \parallel .)$ the
relative entropy.

{\it Remark 1:} \,
The important inequality
$$
\chi(T( {\cal E} )) \leq \chi({\cal E})
$$
follows from the monotonicity of relative entropy which
can be proved for trace preserving, at least 2-positive maps.
Counter examples for
just positive maps seems to be unknown.
There are two recent reviews on the finer
properties of relative entropy, one by Petz \cite{Pe02},
going beyond complete positivity, and one by
Ruskai \cite{Ru02a}.

Let $T$ be a  positive trace preserving map acting on the
density operators of $\cH$. The application of $\rho \to T(\rho)$
generates, letter by letter, a new quantum message belonging
to the new ensemble
\begin{equation} \label{u13a}
T({\cal E}) = \{ T(\rho_1), \dots, T(\rho_m) ; p_1, \dots, p_m\}
\end{equation}
The {\em 1-shot} or {\em Holevo capacity} of a channel $T$
is the number
\begin{equation} \label{u14}
{\bf C}^{(1)}(T) := \max_{{\cal E}} \chi( T({\cal E}) )
\end{equation}
We are aiming at a slightly more delicate expression. To see
its significance we try to perform (\ref{u14}) in two steps.
At first we let ${\cal E}$ run only through the ensembles
with a given average density $\rho$, postponing the search for
the maximum, i.~e.~the Holevo capacity, to a later time.
  In doing so we first define
\begin{equation} \label{u15}
H(T \, ; \rho) := \max_{ \m[{\cal E]} = \rho} \chi( T({\cal E}) )
\end{equation}
so that the maximum is to compute with respect to all ensembles
having $\rho$ as its average density. We get a function, depending
on $\rho$ (and $T$) only. If $T$ is a partial trace onto a
sub-algebra, (\ref{u15})
has been called {\em entropy of $\rho$ with respect to the
sub-algebra} by Connes et al \cite{CNT87}, see also Narnhofer
and Thirring \cite{NT85}. In this sense (\ref{u15})
could be called {\em entropy of $\rho$ with respect to $T$.}

Schumacher and Westmoreland
\cite{SW99} relate (\ref{u15}) to the efficiency   of
quantum channels. They denote the quantity in question
by $\chi^*(\rho)$ and they call an ensemble
saturating (\ref{u15}) an {\em optimal signal ensemble.}

It is important for our purposes to rewrite (\ref{u15})
by the help of (\ref{u13}) as
\begin{equation} \label{u16}
H(T \, ; \rho) = S( T(\rho) )   -  E(T \, ; \rho)
\end{equation}
where $E$ is an entanglement measure given by
\begin{equation} \label{u17}
E(T \, ; \rho) = \min_{ \sum p_k \rho_k = \rho} \sum p_j S(T(\rho_j))
\end{equation}
and the minimum is to compute over all convex decompositions
of $\rho$. If $T$ is a partial trace in a bipartite quantum
system,(\ref{u17}) is the {\em entanglement of formation}
introduced in Bennett et al \cite{BDSW96}, followed by the
remarkable papers of Wootters, \cite{Woo97},  and of
Terhal and Vollbrecht \cite{TV00}.
Other examples are in Benatti et al \cite{BNU96},
\cite{BNU99}. In \cite{BNU02} some
relations between the quantity (\ref{u17}) for quite
different
channel maps are pointed out by the same authors.

It is well known that $E(T \, ; \rho)$ is convex in $\rho$.
Indeed,  $E$ is written as the convex hull of the
function $S(T(\rho))$ in  (\ref{u17}).
The concavity of the entropy allows to perform (\ref{u17})
over the pure decompositions of $\rho$ without changing the
outcome of the minimization. Thus
\begin{equation} \label{u18}
E(T \, ; \rho) = \min \sum p_j S(T(\pi_j)),
\quad \rho = \sum p_j \pi_j
\end{equation}
where all the $\pi_k$ are pure and the minimum is to perform
with respect to all pure decompositions of $\rho$. An ensemble
of pure states saturating (\ref{u18}) will be called an
{\em optimal ensemble.}

One should also notice: The equality between the two
variational problems (\ref{u17}) and (\ref{u18}) is due to
the concavity of $S(T(\rho))$.  If we perform similar
computations, however with a function not being concave,
the two problems are essentially different.
For certain channels, including all rank two and length
two 1-qubit channels, the minimization (\ref{u18})
will be solved in the present paper.

The next section is a short introduction to the roof concept.
We shall discuss how to
estimate (\ref{u18}) from below by convex functions
and from above by roofs.

Next we demonstrate the procedure for the phase-damping
channels, where most things are now well understood.
It follows the computation of concurrences and of $E$
for some rank two channels. We give quite explicit
computations to see the dependence of concurrences,
foliations, and optimal ensembles from the Kraus operators.

A quite interesting observation is the following: Let us
call {\em Kraus module} of $T$ the linear span of the
Kraus operators of $T$. Choi \cite{Ch75} has shown that
irreducibility
of trace preserving cp-maps is a sole property of the Kraus
module. Here we prove for length two and rank two channel
maps that their optimal ensembles
and, hence, their foliations are equal if they belong
to the same Kraus module. It is not probable that this
remarkable feature will survive for more complex channels.
Nevertheless it seems worthwhile to ask for similarities
of channels with identical Kraus modules.

It is a further consequence from our calculations that
the foliations deform continuously by changing
the entries of the Kraus operators for the class of
channels considered. The foliations are more
coarse properties of rank two cp-maps than concurrences.

\section{Roofs}
There are some general features in the optimization problems
we are interested on. They constitute the ground floor for more
refined investigations.

Let us abstract from the specific values given at the pure
states in (\ref{u18}) and let us start with an arbitrary
real valued and continuous function $g(\pi)$ on the set of
pure states. Assume,  we like to extend $g$ to a function
defined for all density matrices. Clearly, there are many
and quite arbitrary solutions for the problem.
Let us denote by $G$ one of these extension.

To place a first restriction,
we require the extension to be ``as linear as possible''.
The requirement can be made precise as following. Let
$\rho$ be a density operator. If there exist a pure
decomposition of $\rho$ such that
\begin{equation} \label{u19}
G(\rho) = \sum p_j g(\pi_j), \quad \rho = \sum p_j \pi_j,
\end{equation}
 we call the decomposition {\em $\rho$-optimal} or
simply {\em optimal} for $G$. We also call a pure ensemble
$$
{\cal E} = \{ \pi_1, \dots, \pi_m ; p_1, \dots, p_m\},
\quad p_j > 0,
$$
$G${\em -optimal} if
$$
G( \sum p_j \pi_j) = \sum p_j G(\pi_j)
$$

A function, $G$, which allows an optimal decomposition
(\ref{u19}) for every state $\rho$, I call a {\em roof}
or, more literally, {\em a roof extension of $g$.}
Roof extensions reflect the convex structure of the
state space and they are, in a well defined way,
``as linear as possible''. It is not easy to gain
good examples of roof extensions in higher
dimensions.  In general, however, there are a lot of them
for a given function $g$.

Let us now consider two further possibilities to extend
$g$ from the pure ones to all states:
We may require the extension to be either concave
or convex. In this spirit we call a function on the state
space a {convex extension of $g$} if the extension is a
convex function. Similar we speak of a {\em concave
extension of $g$} if the extension is concave.

{\bf Lemma 1.} \, {\em Given a convex, a concave,
and a roof extensions of $g$. Then }
\begin{equation} \label{u110a}
G_{{\rm convex}}(\rho) \leq G_{{\rm roof}}(\rho)
\leq G_{{\rm concave}}(\rho)
\end{equation}
{\em for all density operators $\rho$.}

The proof is almost trivial: One chooses a $\rho$-optimal
decomposition (\ref{u19}). Then the very definition of
a convex (concave) function establishes (\ref{u110a}).

It follows that, given $g$, there can be only one convex roof
extension, ``the'' {\em convex roof} determined by $g$.
The family of roof extensions of $g$
has just one member in common
with the family of its convex extensions.
Similarly there is just one {\em concave roof} which
extends $g$.

As a matter of fact \cite{Uh98}, the convex roof with
values $g(\pi)$ at the pure states is nothing
than the solution of the
variational problem which mimics (\ref{u18}).
\begin{equation} \label{u110}
G_{{\rm convex, roof}}(\rho) =
\min \sum p_j g(\pi_j), \quad \rho = \sum p_j \pi_j
\end{equation}
Here one has to run through all pure state
decompositions of $\rho$. We get (\ref{u18})
by setting $g(\pi) = S(T(\pi))$ for pure $\pi$.
On
the other hand, if we take the maximum in (\ref{u110})
instead of the minimum, we obtain the concave roof
extension of $g$.

Let us return for a moment to the more specific of
evaluating (\ref{u18}). To calculate
$E(T \, \rho)$ amounts to construct the convex roof with
the function $g(\pi) = S(T(\pi))$. Looking at the roof
property, there are some typical questions one should
ask. For instance we may start by a set of
parameterized mappings,
$T_s$, and we like to know whether they have, perhaps
for some $\rho$, identical optimal decompositions.
More literally, we ask for a pure decomposition of
$\rho$ which is optimal for every $T_s$ in an
appropriate range of the parameter $s$. In the next
sections we like to convince the reader that this
point of view is quite fruitful. To do so, a
remarkable property of convex and concave roofs
is to explain.

Let $G$ be a convex roof on the density operators of
a Hilbert space of finite dimension $d$. At first we use
convexity: Let us fix a density operator $\rho$.
There is at least one Hermitian operator, say $Y^{\rho}$,
such that for all density operators $\omega$
\begin{equation} \label{u111}
G(\rho) = \T \, Y^{\rho} \rho, \quad
G(\omega) \geq \T \, Y^{\rho} \omega
\end{equation}
is valid. Now let us apply the roof property: There is
a pure decomposition
$$
\rho = \sum p_j \pi_j, \quad p_j > 0, \quad \pi_j \, \, \hbox{pure,}
$$
 which is $G$-optimal. Thus
$$
\sum p_j \T \, Y^{\rho} \pi_j =
\T \, Y^{\rho} \rho = G(\rho) = \sum p_j G(\pi_j)
$$
Because of the inequality (\ref{u111}) this can hold if
and only if
$$
G(\pi_j) = \T \, Y^{\rho} \pi_j, \quad j = 1, 2, \dots
$$
Let us now look at the convex set
\begin{equation} \label{u112}
\{ \, \omega \, : \, \, G(\omega) = \T \, Y^{\rho} \omega \, \}
\end{equation}
By the very construction, $G$ is convexly linear (affine)
if restricted to this set. On the other hand,
it contains $\rho$ and, by the reasoning above, it contains every
pure state which belongs to a $G$-optimal ensemble with
average density $\rho$. Or, in other words, (\ref{u112})
contains all the pure states which can appear
in an optimal decomposition of $\rho$. Let us collect
all that in a theorem, \cite{BNU96}, \cite{Uh98}.

{\bf Theorem 2.} \, {\em Let $g$ be a continuous function
on the
pure states.\\
i)  There exist exactly one convex roof extension,
$G$, of $g$.\\
ii) $G$ can be represented by the optimization
procedure (\ref{u110}).\\
iii) There exist optimal pure  decompositions for every density
operator $\rho$.\\
iv) Given $\rho$, $G$ is convexly linear on the convex hull
of all those pure states which can appear in an optimal
decomposition of $\rho$.}

Let us call the convex hull of all pure states
appearing in all possible optimal pure decompositions of
$\rho$ the {\em optimal convex leaf} of $\rho$. As proved
above, and stated in the theorem, $G$ must be convexly linear
on every optimal convex leaf.

{\it Remark 2.} The remark concerns property iii).
The compact convex set of density operators enjoys a
peculiarity: The set of their extremal points, i.~e.~the set of
pure states, is a compact one. This
allows to prove the existence of
optimal decompositions (\ref{u19}) by the
continuity of $g$. Then, by a theorem due to
Caratheodory, one deduces the existence of optimal
decompositions with a length not exceeding $d^2$,
$d$ the dimension of the Hilbert space which carries
the density operators. It should be noticed that the
compactness of the extremal points is an extra property.
Counter examples are by no means exotic as seen
by the set of trace preserving cp-maps.
To get the conclusion iii) it suffices for $g$ to be
continuously extendable to the closure of
the set of extremal states.

 {\it Remark 3.} If $G$ is a convex roof and $G$ is the sum
  of two convex functions $G = G_1 + G_2$, then $G_1$ and $G_2$
 are convex roofs, and every $G$-optimal ensemble is optimal
 for $G_1$ and $G_2$.

 We need one more definition: We call a convex roof, $G$,
 {\em flat}
if it allows for every $\rho$ an optimal pure decomposition
$$
\rho = \sum p_j \pi_j, \quad G(\rho) = \sum
p_j G(\pi_j)
$$
such that
$$
G(\pi_1) = G(\pi_2) = \dots = G(\pi_j) = \dots
$$
If this takes place, every $\rho$ is contained in a convex
subset, generated by pure states, on which
the roof is not only convexly linear but even constant.
The merit of flat roof, say $G$, is in the nice property
that every function of $G$, say $f(G)$,  is again a roof.

As a matter or fact, the convex roofs we are considering
in the following enjoy even a stronger property: They are
constant on the convex leaf of every $\rho$, i.~e. $g$ is
constant on the pure states of {\em every} pure ensemble
which is optimal for $G$.

\section{Phase damping channels}
Let us consider some particularly simple examples
of 1-qubit channels, the family of
phase-damping channels. to see what is going on in applying
concurrences, \cite{Woo97},
to 1-qubit channels according to \cite{Uh01}. For the most
symmetric channel of the family, with $z=0$ in (\ref{u21}),
the theorem below and the insight into the foliation
of the state space are due to Levitin \cite{Le94}.

Let $|z| < 1$ be a complex number.
 Define the map $T_z$ by
\begin{equation} \label{u21}
X = \pmatrix{x_{00} & x_{01} \cr x_{10} & x_{11}}
\, \mapsto \,
T_z(X) = \pmatrix{x_{00} & z x_{01} \cr z^* x_{10} & x_{11}}
\end{equation}
The application of such a map does not change the pure states
$|0 \rangle\langle 0|$ and $|1 \rangle\langle 1|$, and there are
no other trace preserving and completely positive 1-qubit maps
with this property than those given by (\ref{u21}).

Before starting the calculation let us have a look on a
bundle of parallel lines which foliates the state space.
Geometrically, the 1-qubit state space can be represented
by the Bloch ball. The Bloch ball is the unit ball sitting in
the ``Bloch space'', that is in the
real Euclidean 3-space of all Hermitian matrices of trace one.
The ``Pauli coordinates'', $x_j$, of a matrix
are read off from
$$
X = {1 \over 2} \, (x_0 \1 + x_1 \sigma_1 + x_2 \sigma_2 + x_3 \sigma_3)
$$
With $x_0 = 1$ the real coordinates $x_1$, $x_2$, $x_3$,
parameterize the Bloch space, and in this context they
are referred to as ``Bloch coordinates''. Finally, the Bloch ball
is the unit ball with respect to the Bloch coordinates.

Let us return to the phase-damping channels.
The line $x_1 = x_2 = 0$, i.~e.~the $x_3$-axis,
 remains point-wise fixed under the
mappings (\ref{u21}). On the intersection of the line with
the Bloch ball $E(T_z ; \rho)$ must be zero. One aim is
to show that $E(T_z ; \rho)$ is constant on the intersection
of the Bloch ball with every line which is parallel to
the line $x_1 = x_2 = 0$. These lines are given by fixing the
values of $x_1$ and $x_2$ and letting $x_3$ free. Equivalently,
such a line can be given by fixing
$x_{01}$ in (\ref{u21}). Now we can write down a very simple
convex roof, the restriction of the function
$$
X \mapsto  \sqrt{x_1^2 + x_2^2} \equiv 2 \, |x_{01}|
$$
onto the state space. Indeed, this function is convex on the
Bloch space. It is even a semi-norm there. And it is, trivially,
a flat roof: It is constant on every line
parallel to the $x_3$-axis  of the Bloch space.
The intersection of a given line with the Bloch ball
is either empty, or touching the ball in just one point, or
is a line segment with two pure states as end points.
In the latter case, the pure states of the segment are
\begin{equation} \label{ext1}
\pmatrix{1-p & x_{01} \cr x_{10} & p} , \, \,
\pmatrix{p & x_{01} \cr x_{10} & 1-p}, \quad
p(1-p) = |x_{01}|^2
\end{equation}
with $2|x_{01}| < 1$. For $2|x_{01}| = 1$ the line touches
the Bloch ball at one pure state.

With a flat roof one can build other roofs just by taking
a function of it: If $f$ is a real function on the
unit interval, $f(|x_{01}|)$ is again a flat roof. By an
appropriate choice of $f$ we shall find the form of
$E(T_z ; \rho)$.

To do so we need to compute the determinant
$$ \det T_z(X) = x_{00} x_{11} - zz^* x_{01} x_{10}
= \det X + (1 - zz^*) x_{01} x_{10}
$$
Taking $X$ pure, say $x_{jk} = a_j a_k^*$, only the second
term is different from zero and we remain with
\begin{equation} \label{u22}
\det T_z(X) = (1 - |z|^2) |a_0 a_1|^2,  \quad x_{jk} = a_j a_k^*
\end{equation}
Using ideas from \cite{Woo97} and \cite{Uh00c} we define
the {\em concurrence} of $T_z$ for Hermitian $X$ by
\begin{equation} \label{u22a}
C( T_z \, ; X) := \sqrt{(1 - |z|^2) \, (x_1^2 + x_2^2) }
\end{equation}
The definition differs from the one used in \cite{Uh01}
by a factor
two. It influences, here and later on, the appearance of
some equations.
The concurrence is a semi-norm on the Bloch ball, and,
if restricted to the Bloch ball, it is the unique convex
roof satisfying
\begin{equation} \label{u22b}
C( T_z \, ; \pi) = 2 \sqrt{\det T_z(\pi)}, \quad \pi \,
\hbox{pure}
\end{equation}
Following again \cite{BDSW96} and \cite{Woo97} we introduce
$$
h(x) = - x \log x - (1-x) \log (1-x)
$$
and, using ad hoc notations,
$$
h_1(x) = h({1+x \over 2})
$$
and
$$
h_2(x) = h_1(y), \quad y = \sqrt{1 - x^2}
$$
{\bf Theorem 3.} \, {\em For all $|z| < 1$ and all
density operators $\rho$
\begin{equation} \label{u22c}
E( T_z \, ; \rho) = h_2( \, C( T_z \, ; \rho) \, )
\end{equation}
holds. It is $E(\rho_1) = E(\rho_2)$ for two density operators,
$\rho_1$ and $\rho_2$, if and only if they have equal distances
to the $x_3$-axis of the Bloch space. The pairs of optimal
pure states are given by (\ref{ext1}).}

Proof: We already know that (\ref{u22c}) is a roof with the
desired values at the pure states. We only have to show that
it is convex. Then, by the uniqueness theorem, we are done.
One calculates the first and the second derivative of $h_2$,
assuming $\log \equiv \ln$. At first we get
$$
h'_2(x) = {x \over 2y} \ln {1+y \over 1-y}
$$
For $x \geq 0$ and $y \geq 0$ we find $h'_2 \geq 0$. One
further obtains
$$
h''_2(x) = {1 \over 3} + {y^3 \over 5} + {y^5 \over 7}
+ {y^7 \over 9} + \dots
$$
and $h''_2 \geq 0$ proves the convexity of $h_2$. Let $C(\rho)$
be a convex function on the state space ( - or on another
convex set - ) with values between 0 and 1.
 Let us denote by a dot the differentiation
of $C$ in an arbitrary direction. Then
$$
h_2(C)^{\cdot \cdot} = h''_2(C) \dot C \dot C + h'_2(C) \ddot C
$$
By the convexity of $C$ one gets $\ddot C \geq 0$, and we have
seen $h' \geq 0$ on the unit interval. Thus, the second term
is not negative. As we know $h'' \geq 0$, we have shown
the convexity of the function  (\ref{u22c}), and we done.

{\bf Lemma 4.} \,{\em Let $C$ be a convex function
with values in the unit interval, defined on a
finite dimensional convex set. Then $h_2(C)$
is convex.}

It is remarkable that the whole set of phase damping
channels induces a single foliation of the state space:
The foliation is a property of the Kraus module belonging
to the channels. The single foliation forces the
concurrences to differ by a factor only if $z$ is
changing.

Now we have to add a further structural element as a guide
in treating more general 1-qubit channels.
What we have in mind is, up to a contracting factor,
a reflection on the plane $z_3 = 0$. In Bloch space
a reflection is not a proper rotation. Its functional determinant
must be negative, enforcing its implementation by an anti-linear
operator in the 1-qubit space. The anti-linearity is unavoidable.

Let us define an anti-linear operator $\vartheta_z$ by
\begin{equation} \label{u23}
\vartheta_z \, \pmatrix{a_0 \cr a_1} =
\sqrt{1 - zz^*} \, \pmatrix{a_1^* \cr a_0^*}
\end{equation}
The operator is an Hermitian one, i.~e.~it satisfies
$$
\langle \phi_1 | \vartheta_z | \phi_2 \rangle =
\langle \phi_2 | \vartheta_z | \phi_1 \rangle
$$
for all pairs of vectors. Being anti-linear,
$\vartheta$ acts to the right,
but not to the left. With two arbitrary vectors,
$$
|a\rangle = \pmatrix{a_0 \cr a_1} , \quad
|b\rangle = \pmatrix{b_0 \cr b_1} ,
$$
we get the following relation.
\begin{equation} \label{u23a}
\langle a| \vartheta_z |b \rangle = \sqrt{1 - zz^*} \,
(a_0 b_1 + a_1 b_0)^*
\end{equation}
In combination with (\ref{u22}) we obtain the equation
\begin{equation} \label{u24}
4 \det T_z( |a \rangle\langle a| ) =
| \, \langle a| \vartheta_z |a \rangle \, |^2
\end{equation}
which, together with (\ref{u22b}), can be written
\begin{equation} \label{u24a}
C( T_z \, ;  |a \rangle\langle a| ) =
| \, \langle a| \vartheta_z |a \rangle \, |
\end{equation}
Because the concurrence is a convex roof, we can now return
to one of the properties of such function:
\begin{equation} \label{u25}
C( T_z \, ; \rho ) = \min \sum p_j
| \, \langle \phi_j| \vartheta_z |\phi_j \rangle \, |
\end{equation}
where the minimum runs through all possible ways of representing
$\rho$ as a convex combination
$$
\rho = \sum p_j |\phi_j \rangle\langle \phi_j|
$$
The minimum will be attained by choosing the decomposition
with a pair (\ref{ext1}) of pure states. With
the same optimal decomposition of $\rho$ the optimization
problem for $E(T_z \, ; \rho)$ can be saturated. (The
diagonal entries of $\rho$ must 1/2.)

{\it Remark  4.} \, Let us assume we like to solve for the
phase-damping channels the variational problems (\ref{u17}),
but we like to replace the minimization by the maximization,
resulting in concave roofs. Now the foliation
for the maximization is given by the intersection of
the planes perpendicular to the $x_3$-axis with the Bloch ball.
The foliation is the same for all $z$. The
intersection of a plane with a ball contains a whole circle
of extremal states. Hence there are very many different optimal
decompositions for given mixed $\rho$.

\section{Concurrence}
Let us now discuss some generalities for rank two channels
and let us see, where the difficulties are. A quantum channel,
$T$, is of rank $k$ if the maximal rank of all pictures,
$T(\rho)$, is $k$.

Let $T$ be of rank two. We assume in addition that $T$ maps
into a 1-qubit space. Then the determinant of $T(\rho)$
is the product of the eigenvalues of $T(\rho)$.
We define the {\em concurrence,}
$C(T \, ; \rho)$, of $T$ to be the convex roof  which
attains at pure states the values
\begin{equation} \label{u31}
C( T \, ; \pi) = 2 \sqrt{\det T_z(\pi)}, \quad \pi \,
\hbox{pure}
\end{equation}
completely similar to (\ref{u22b}).
Being convex with
values between 0 and 1, we can literally repeat the
construction  as in (\ref{u22c}). The result is a convex
function which coincides at pure states with $E$. But
in general we
do not know, wether (\ref{u31}) is a {\em flat} roof.
Hence, we only can conclude
\begin{equation} \label{u32}
E( T \, ; \rho) \geq h_2( \, C( T \, ; \rho) \, )
\end{equation}
If, however, (\ref{u31}) is a flat roof, then the right hand
side of (\ref{u31}) is a flat convex roof and we have
two convex roofs agreeing on pure states. Then equality
must hold.

{\bf Lemma 5.} {\em If $C$ as given by (\ref{u31}) is a
convex roof, then we have}
\begin{equation} \label{u32a}
C \, \, \hbox{flat} \, \Rightarrow \, \,
E( T \, ; \rho) = h_2( \, C( T \, ; \rho) \, )
\end{equation}

It is useful to know whether the concurrence of a channel
map $T$ is a flat roof. The following theorem gives a whole
class of them.

{\bf Theorem 6.} \, {\em Let $\vartheta$ be an anti-linear
hermitian
operator and define $C_{\vartheta}$ as the convex roof
extension of
\begin{equation} \label{u33}
C_{\vartheta}( |\phi \rangle\langle \phi| ) =
| \, \langle \phi | \vartheta | \phi \rangle  \,|
\end{equation}
for all pure states $\pi$. Then the convex roof $C_{\vartheta}$
is flat. It is
\begin{equation} \label{u34}
C_{\vartheta}( \rho ) =
\max \{ \, 0 , \lambda_1 - \sum_{j > 1} \lambda_j \, \}
\end{equation}
where $\lambda_1 \geq \lambda_2 \geq \dots$ are the ordered
eigenvalues of}
$$
( \, \rho^{1/2} \vartheta \rho \vartheta \rho^{1/2} \,)^{1/2}
$$

{\bf Corollary 7.} \, If there exists an anti-linear hermitian
$\vartheta$ such that
\begin{equation} \label{u35}
4 \det T(\pi) = \T \, \pi \vartheta \pi \vartheta, \quad
\pi \, \hbox{pure}
\end{equation}
is valid for all pure density operators then
\begin{equation} \label{u36}
C_{\vartheta}(\rho) = C( T \, ; \rho)
\end{equation}
is a flat roof and (\ref{u32a}) is valid.

This theorem is proved in \cite{Uh00c}. It provides a certain
application of the methods of Wootters and others,
see Wootters \cite{Woo01}.
As already mentioned, with $T$ the partial trace, $E(T \, ; \rho)$
is the entanglement of formation. In the 2-qubit system
$\vartheta$ is the Hill-Wootters conjugation.

For 1-qubit maps (\ref{u34}) reads $\lambda_1 - \lambda_2$
and one can become more explicit. Abbreviate
$$
\xi = ( \rho_1^{1/2} \rho_2 \rho_1^{1/2})^{1/2}
$$
Taking the trace of the characteristic equation of
$\xi$ results in
$$
\T \xi^2 + 2 \det \xi =  ( \T \xi )^2
$$
and the squared sum of the eigenvalues of $\xi$ becomes
$$
(\lambda_1 + \lambda_2)^2 =
\T (\rho_1 \rho_2) + 2 \det \xi
$$
Combined with the relation
$$
(\lambda_1 - \lambda_2)^2 = (\lambda_1 + \lambda_2)^2 - 4 \det \xi
$$
it yields
$$
(\lambda_1 - \lambda_2)^2 =
\T (\rho_1 \rho_2) - 2 \sqrt{ (\det \rho_1) (\det \rho_2)}
$$
Substituting $\rho_1 = \rho$ and $\rho_2 = \vartheta \rho \vartheta$
provides, \cite{Uh00c},
\begin{equation} \label{u37}
C_{\vartheta}(\rho)^2 = \T \, (\rho \vartheta \rho \vartheta )
- 2 (\det \rho ) \, \det \sqrt{\vartheta^2}
\end{equation}

\section{1-qubit channels of length two}
In this section we like to show that the corollary to the
preceding theorem applies to 1-qubit channels of length
two. As we shall see, the existence of an anti-linear hermitian
$\theta$ fulfilling (\ref{u35}) does not depend on
trace preserving.

Let $A$ and $B$ be two linear independent operators on a
2-dimensional Hilbert space and
$$
A = \pmatrix{ a_{00} & a_{01} \cr a_{10} & a_{11} },
\quad
B = \pmatrix{ b_{00} & b_{01} \cr b_{10} & b_{11} }
$$
their matrix representations with respect to a reference
basis.
\begin{equation} \label{u41}
T(X) = A X A^{\dag} + B X B^{\dag}
\end{equation}
is a completely positive map of length two. The following has
been proved in \cite{Uh01} by straight forward computation.

{\bf Theorem 8.} \, {\em
There is an anti-linear hermitian operator
$\vartheta_{A,B}$ such that for all pure states
$$
4 \det T( |a \rangle\langle a| ) =
| \, \langle a| \vartheta_{A,B} |b \rangle \, |^2
$$
where $T$ is given by (\ref{u41}).}

If such an operator exists, it is determined
by $T$ up to a phase factor only. The ambiguity is
a natural one due to a geometric phase.

To describe $\vartheta$ one can introduce its matrix representation
\begin{equation} \label{u42}
\vartheta_{A,B} \pmatrix{a_0 \cr a_1} = \pmatrix{
\alpha_{00} a_0^*  + \alpha_{01} a_1^* \cr
\alpha_{10} a_0^* + \alpha_{11} a_1^* }
\end{equation}
and express the matrix elements by those of $A$ and $B$.
\begin{equation} \label{u43}
\alpha_{00} = 2 (b_{10} a_{00} - a_{10} b_{00})^*,
\quad
\alpha_{11} = 2 (a_{01} b_{11} - b_{01} a_{11} )^*,
\end{equation}
\begin{equation} \label{u44}
\alpha_{01} = \alpha_{10} =
(a_{00}  b_{11} - a_{11}  b_{00} + a_{01} b_{10} - a_{10}  b_{01} )^*
\end{equation}
Up to a factor, due to another normalization of the
concurrence, this is agreement with \cite{Uh01}.  Clearly,
$\vartheta$ must be skew symmetric in
the matrix entries of the Krause operators,
$$
\vartheta_{A,B} + \vartheta_{B,A} =0 .
$$
Comparing the equations above with phase-damping
channels one gets $\vartheta_{A,B} = -\vartheta_z$.

One may expect a more transparent
representation than (\ref{u43}) and (\ref{u44}).
This is possible by the spin-flip operator $\theta_f$,
the ``fermion conjugation'' for one
qubit,
$$
\theta_f \pmatrix{a_0 \cr a_1} = \pmatrix{a_1^* \cr - a_0^*}
$$
After calculating
$$
A^{\dag} \theta_f B =
\pmatrix{(b_{10} a_{00} - a_{10} b_{00})^*
& (a_{00} b_{11} - a_{10} b_{01})^* \cr
(a_{01} b_{10} - a_{11} b_{00})^* &
(a_{01} b_{11} - a_{11} b_{01})^*  }
$$
and comparing that expression with (\ref{u43}) and (\ref{u44}),
we get
\begin{equation} \label{u45}
\vartheta_{A,B} = A^{\dag} \theta_f B - B^{\dag} \theta_f A
\end{equation}
Remember $\theta_f^{\dag} = - \theta_f = \theta_f^{-1}$ to see
that (\ref{u45}) is an anti-linear hermitian operator -- as it
should be. Now assume a transformation of
the Kraus operators according to
\begin{equation} \label{45a}
A, \, B \, \longrightarrow \, C_1 A C_2, \, C_1 B C_2
\end{equation}
Because of
$$
C_1^{\dag} \theta_f  C_1 =  ( \det C_1 )^* \theta_f
$$
the anti-linear operator must change as follows:
\begin{equation} \label{45b}
\vartheta_{A,B} \, \longrightarrow \, ( \det C_1 )^* \,
C_2^{\dag} \vartheta_{A,B} C_2
\end{equation}
That it is of value to classify super-operators up to a
transformation (\ref{45a}) is also seen by the results
of Verstraete and Verschelde, \cite{VV02}.

Let $T'$ be another length two cp-map with Kraus coefficients
$A'$ and $B'$, and let us assume a linear dependence
\begin{equation} \label{u46}
A' = \mu_{00} A + \mu_{01} B, \quad
B' = \mu_{10} A + \mu_{11} B
\end{equation}
By the help of (\ref{u45}), or by observing that (\ref{u43})
and (\ref{u44}) can be expressed by determinants in the
coefficients of the Kraus operators, one can reproduce
the relation
\begin{equation} \label{u47}
\vartheta_{A',B'} =
(\mu_{00} \mu_{11} - \mu_{01} \mu_{10})^* \vartheta_{A,B}
\end{equation}
Why is this interesting? It shows that our procedure associates,
up to a scalar factor, to every pair of operators, chosen from
the linear span of $A$ and $B$, the same anti-linear operator.
With other words, to every 2-dimensional Kraus module
 a 1-dimensional linear space of anti-linear
hermitian operators is attached.

{\it Remark 5.} Regard the Kraus modules for the 1-qubit
channels as the points of the second Grassmann
manifold of the space of linear operators. By attaching the
multiples of $\vartheta_{A,B}$ to the corresponding points
one gets the line bundle. It is dual to the determinant bundle
as one can deduce from what follows. For the time being,
we shall not follow further this way.

An observation, related to (\ref{u47}), is
\begin{equation} \label{u49}
A' \otimes B' - B' \otimes A'  = (\mu_{00} \mu_{11} - \mu_{01} \mu_{10})
\, (A \otimes B - B \otimes A )
\end{equation}
If we apply the operator identity (\ref{u49}) to a $|aa\rangle$,
we get an anti-symmetric 2-qubit vector. There is, up to
a factor, only one such vector and the yet unknown factor must
transform as in (\ref{u49}). Performing the calculations one
gets
\begin{equation} \label{u410}
\, (A \otimes B - B \otimes A ) \, |aa\rangle = {1 \over 2} \,
  \langle a| \vartheta_{A,B} |a \rangle^* \,
( |01\rangle - |10\rangle)
\end{equation}
Remember that for a single channel, $T$, only the absolute
value of the expectation value
at the right hand side is relevant.

Two channels may be called ``unitary equivalent'' if
$$
T'(\rho) = U_1 \, T( U_2^{-1} \rho U_2 ) \, U_1^{-1}
$$
with two unitaries $U_1$ and $U_2$. For trace preserving channels
we can assume
\begin{equation} \label{u50}
A = \pmatrix{ a_{00} & 0 \cr 0 & a_{11} },
\quad
B = \pmatrix{ 0 & b_{01} \cr b_{10} & 0 }
\end{equation}
up to unitary equivalence, \cite{RSW00} . Then $\vartheta$
becomes diagonal,
\begin{equation} \label{u51}
\vartheta_{A,B} \pmatrix{a_0 \cr a_1} = 2 \, \pmatrix{
(b_{10} a_{00} a_0 )^* \cr
(- b_{01} a_{11} a_1 )^* }
\end{equation}
Abbreviating
$$
y_0 = 2 b_{10} a_{00}, \quad y_1 =  2 b_{01} a_{11}
$$
we can write
$$
\vartheta \rho \vartheta = \pmatrix{ \rho_{00} y_0 y_0^* &
- \rho_{10} y_0 y_1^* \cr - \rho_{01} y_1 y_0^* & \rho_{00} y_1 y_1^*
}
$$
It follows
$$
\T \,\rho \vartheta \rho \vartheta =
\rho_{00}^2 y_0 y_0^* -
\rho_{01}^2 y_0 y_1^* -
\rho_{10}^2 y_1 y_0^* +
\rho_{00}^2 y_1 y_1^*
$$
On the other hand,
$$
2  \det \rho \, \det \sqrt{\vartheta^2}  = 2 \, |y_0 y_1| \,
(\rho_{00} \rho_{11} - \rho_{01} \rho_{10})
$$
Pasting all things together, we get from (\ref{u37})
$$
C^2 = (\rho_{00} |y_0| - \rho_{11} |y_1|)^2 + 2 |y_0 y_1|
\rho_{01} \rho_{10} - \rho_{01}^2 y_0 y_1^* - \rho_{10}^2 y_0^* y_1
$$
We now choose the square roots of $y_0 y_1^*$ and $y_0^* y_1$
such that their product becomes positive. Then we can write
the remaining terms above as follows:
$$
- ( \rho_{01} \sqrt{y_0 y_1^*} - \rho_{10} \sqrt{y_0^* y_1})^2
$$
so that, finally, we see:
\begin{equation} \label{u52}
C(T \, \rho)^2 = 4 L_1(\rho)^2 + 4 L_2(\rho)^2
\end{equation}
were $L_1$ and $L_2$ are real valued and linear in the entries
of $\rho$,
\begin{equation} \label{u53}
L_1(\rho) = \rho_{00} |b_{10} a_{00}| - \rho_{11} |b_{01} a_{11}|
\end{equation}
\begin{equation} \label{u54}
L_2(\rho) = i ( \, \rho_{01} \sqrt{b_{10} a_{00} b_{01}^* a_{11}^*}
- \rho_{10} \sqrt{b_{10}^* a_{00}^* b_{01} a_{11}} \, )
\end{equation}
and we have to choose the signs of the roots according to
\begin{equation} \label{55}
\sqrt{b_{10} a_{00} b_{01}^* a_{11}^*} \,
\sqrt{b_{10}^* a_{00}^* b_{01} a_{11}} \geq 0
\end{equation}
The result compares well with the more symmetrical case of
the phase-damping channels: $C$ is the square root of a
positive semi-definite quadratical form. Geometrically,
the points of constant concurrence and, hence, of $E$ are
ellipse-based cylinders.

In the non-degenerate case none of the two linear forms vanish
identically. The foliation of the state space at which the
concurrence and also $E(T \, ; \rho)$ remain constant are given
by the straight lines which are the intersection of the two
families of planes $L_1 =$ constant, $L_2 =$ constant', in the
Bloch space. There is just one straight line at which the concurrence
is zero. It goes through the two pure states which are mapped onto
pure states by $T$. Up to normalization these two pure states
belong to the vectors
$$
\pmatrix{a_0 \cr a_1}, \quad b_{10} a_{00} a_0^2 =  b_{01} a_{11} a_1^2
$$
They are mapped by $T$ to vector states of the form
$$
\pmatrix{a'_0 \cr a'_1}, \quad b_{10} a_{11} (a'_0)^2 =
b_{01} a_{00} (a'_1)^2 = 0
$$

Let us now shortly look at the {\em degenerate case} in
which $b_{01} b_{10}$ is zero in (\ref{u50}).
The amplitude-damping channels are well known examples.
They can be defined by the Kraus operators
\begin{equation} \label{u56}
A = \pmatrix{ 1 & 0 \cr 0 & \sqrt{p} },
\quad
B = \pmatrix{ 0 & \sqrt{1-p} \cr 0 & 0 }
\end{equation}
with $0 < p < 1$.
The action of $T$ is described by
$$
\pmatrix{\rho_{00} & \rho_{01} \cr \rho_{10} & \rho_{11} }
\mapsto
\pmatrix{\rho_{00} + (1-p) \rho_{11} & \sqrt{1-p} \rho_{01}
\cr \sqrt{1-p} \rho_{10} & p \rho_{11} }
$$
(\ref{u52}) becomes
\begin{equation} \label{u57}
C(T \, ; \rho) = 2 \sqrt{p(1-p)} \, \rho_{11}
\end{equation}
The two families of planes degenerate to one family, the planes
perpendicular to the 3-axis of the Bloch space. The foliation
dictates the behavior of $C(T \, ; \rho)$ and $E(T \, ; \rho)$.

One observes that, given $\rho_{11}$, the entropy $S(T(\rho))$
becomes maximal if the off diagonal entries of $\rho$ vanish.
Therefore, if $\rho'$ is the diagonal part of $\rho$, We get
$$
E(T \, ; \rho) = E(T \, ; \rho'), \quad H(T \, ; \rho)
\leq H(T \, ; \rho')
$$
and to obtain the Holevo capacity it suffices to consider diagonal
density operators only:
$$
{\bf C}^{(1)} = \max H(T \, ; \rho'), \quad \rho' \, \, \hbox{diagonal}
$$
Writing $r$ for $\rho_{11}$, such that $0 \leq r \leq 1$, we can rewrite
the capacity as follows:
\begin{equation} \label{u58}
{\bf C}^{(1)}(T) = \max_{0 \leq r \leq 1}  [ h(pr) -
h({1 - \sqrt{1 - 4p(1-p) r^2} \over 2}) ]
\end{equation}
Because $H(T \, ; \rho)$ is a concave function, (\ref{u58})
is a concave function of $r$ and, obviously, not degenerate.
Therefore, for any given $0 < p < 1$, there is exactly one
value $r_0$ of $r$
at which the maximum in (\ref{u58}) is attained.

email: armin.uhlmann@itp.uni-leipzig.de
\end{document}